\documentclass[twocolumn,aps,graphicx]{revtex4}

\usepackage[dvips]{graphicx}
\newcommand{\mypicwidth}{0.6\columnwidth}
\newcommand{\myext}{ps}
\newcommand{\n}[2]{n_{#1}^{#2}}
\newcommand{\nv}[2]{{\bf n}_{{\rm v}#1}^{#2}}

\newcommand{\w}[2]{w_{#1}^{#2}}
\newcommand{\wv}[2]{{\bf w}_{{{\rm v}#1}}^{#2}}

\newcommand{\wm}[2]{{\bf \hat{w}}_{{{\rm m}#1}}^{#2}}
\newcommand{\mymat}[1]{{\hat{\mathbf{#1}}}}

\newcommand{\nmtl}[2]{{\mymat{n}_{{\rm m}#1}}^{#2}}

\newcommand{\colloid}{C}

\newcommand{\polymer}{P}

\newcommand{\rs}{\rho_{S}^{\ast}}
\newcommand{\rsr}{\rho_{S}^{\ast r}}

\newcommand{\ep}{\eta_{\polymer}}
\newcommand{\ec}{\eta_{\colloid}}

\newcommand{\Fexc}{F_{\rm exc}}

\newcommand{\kb}{k_{\rm B}}
\newcommand{\kt}{k_{\rm B} T}

\newcommand{\phifmt}[1]{\Phi_{#1}}

\newcommand{\rv}{{\bf r}}

\newcommand{\fumindex}{\nu}

\begin{document}
\author{Matthias Schmidt\footnote{
Permanent address: Institut f{\"u}r Theoretische Physik II,
  Heinrich-Heine-Universit{\"a}t D{\"u}sseldorf,
  Universit{\"a}tsstra{\ss}e 1, D-40225 D{\"u}sseldorf, Germany.
} and Alan R. Denton}
\title{
Demixing of colloid-polymer mixtures in poor solvents }
\date{3 October 2001}
\date{6 October 2001}
\date{26 October 2001}
\date{10 November 2001}
\date{4 December 2001}
\date{11 December 2001}
\date{6 March 2002}
\date{5 April 2002}
\affiliation{
Department of Physics, North Dakota State University, Fargo, ND
58105-5566, USA}

\begin{abstract}
The influence of poor solvent quality on fluid demixing of a 
model mixture of colloids and nonadsorbing polymers is investigated
using density functional theory. The colloidal particles are
modelled as hard spheres and the polymer coils as effective
interpenetrating spheres that have hard interactions with the
colloids. The solvent is modelled as a two-component mixture of a
primary solvent, regarded as a background theta-solvent for the 
polymer, and a cosolvent of point particles that are excluded from 
both colloids and polymers. Cosolvent exclusion favors overlap of
polymers, mimicking the effect of a poor solvent by inducing an
effective attraction between polymers. For this model, a
geometry-based density functional theory is derived and applied to
bulk fluid phase behavior. With increasing cosolvent concentration
(worsening solvent quality), the predicted colloid-polymer binodal
shifts to lower colloid concentrations, promoting demixing. 
For sufficiently poor solvent, a reentrant demixing transition 
is predicted at low colloid concentrations.
\end{abstract}

\maketitle

\newpage
\section{Introduction}
Solvents play a crucial role in the thermodynamic behavior 
of macromolecular solutions.  Over the past half-century, 
effects of solvent quality on the physical properties 
of polymer solutions have been extensively 
studied~\cite{flory69,deGennes79}.
Polymer-solvent and solvent-solvent interactions were first 
incorporated into the classic Flory-Huggins mean-field theory 
of polymer solutions~\cite{flory71}.
Subsequently, excluded-volume interactions between polymer 
segments were identified as the key determinants of
solvent quality. Polymer segments sterically repel one another 
in a good solvent, attract in a poor solvent, and behave 
as though ideal (noninteracting) in a theta-solvent.  
Interactions between polymer segments strongly influence 
chain conformations and, in turn, phase separation and other 
macroscopic phenomena.

Compared to solvent effects in pure polymer solutions, much less is 
known about the role of solvent quality in colloid-polymer mixtures.
The simplest and most widely-studied theoretical model of 
colloid-polymer mixtures is the Asakura-Oosawa (AO)
model~\cite{asakura54,vrij76}. This treats the colloids as hard
spheres and the polymers as effective spheres that are mutually
noninteracting but have hard interactions with the colloids. 
The thermodynamic phase diagram of the AO model has been 
mapped out by thermodynamic perturbation theory~\cite{gast83}, 
free volume theory~\cite{lekkerkerker92}, density functional (DF) 
theory~\cite{schmidt00cip}, and Monte Carlo simulation~\cite{dijkstra99}.
By assuming ideal polymers, however, the AO model is implicitly 
limited to theta-solvents. 
Recently, by incorporating polymer-polymer repulsion into the AO
model, the influence of a good solvent on phase behavior has been
explored via perturbation theory~\cite{warren95} and DF
theory~\cite{schmidt02intpol}. All of these studies assume an 
effective penetrable-sphere model for the polymer coils, which 
is supported by explicit Monte Carlo simulations of interacting
segmented-chain polymers~\cite{louis00,bolhuis01jcp,bolhuis01pre}.
An alternative, more microscopic, theoretical approach is the 
PRISM integral-equation theory~\cite{ramakrishnan02}, 
which models polymers on the segment level.

The purpose of the present paper is to investigate the effect of 
a {\it poor} solvent on the bulk phase behavior of colloid-polymer
mixtures. To this end, we consider a variation of the AO model
that explicitly includes the solvent as a distinct component.
Specifically, the solvent is treated as a binary mixture of a
primary solvent, which alone acts as a theta-solvent for the
polymer, and a cosolvent, which acts as a poor solvent for the
polymer. The primary solvent is regarded as a homogeneous
background that freely penetrates the polymer, but is excluded by
the colloids. The cosolvent is modelled simply as an ideal gas of
point-like particles that penetrate neither colloids nor polymers.

In the absence of colloids, the polymer-cosolvent subsystem is the
Widom-Rowlinson (WR) model of a binary 
mixture~\cite{widom70,rowlinson82}, in which
particles of unlike species interact with hard cores and particles
of like species are noninteracting. The WR model can be shown 
to be equivalent to a one-component system of penetrable spheres
that interact via a many-body interaction potential, proportional 
to the cosolvent pressure and the volume covered by the spheres
(with overlapping portions counted only once). 
Hence, in the polymer-cosolvent subsystem, the volume occupied 
by the polymer spheres costs interaction energy, inducing
an effective attraction between polymers reminiscent of that 
caused by a poor solvent.  By varying cosolvent concentration, 
the solvent quality can be tuned.  Here we investigate whether 
and how added hard colloidal spheres mix with such effectively 
interacting polymers.

In Sec.\ \ref{SECmodels}, we define more explicitly the model
colloid-polymer-cosolvent mixture.  In Sec.~\ref{SECtheory}, we
develop a general geometry-based DF theory, which may be applied 
to both homogeneous and inhomogeneous states of the model system.
The general theory provides the foundation for an application 
to bulk phase behavior in Sec.~\ref{SECresults}.
Readers who are interested only in bulk properties may wish to skip 
Sec.~\ref{SECtheory} and turn directly to Sec.~\ref{SECresults}.
We finish with concluding remarks in Sec.~\ref{SECdiscussion}.

\section{The Model}
\label{SECmodels} 
\begin{figure}
  \begin{center}
    \includegraphics[width=\columnwidth,angle=0]{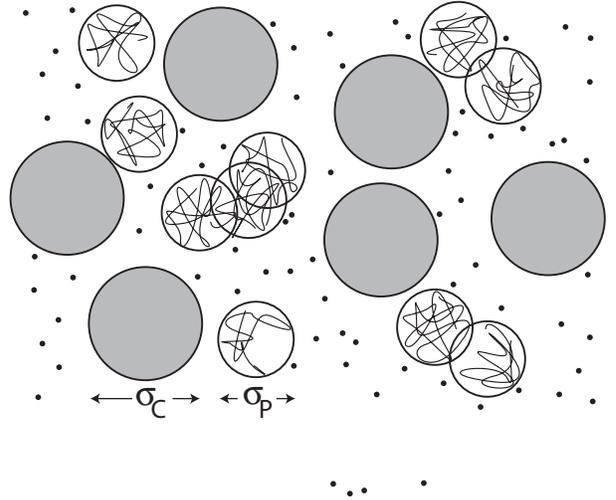}
    \caption{Model ternary mixture of colloidal hard spheres of diameter
    $\sigma_C$, polymer effective spheres of diameter $\sigma_P$, and 
    point-like solvent particles. } \label{FIGmodel} 
  \end{center}
\end{figure}

We consider a ternary mixture of colloidal hard
spheres (species $C$) of radius $R_C$, globular polymers (species
$P$) of radius $R_P$, and point-like cosolvent particles (species
$S$), as illustrated in Fig.~\ref{FIGmodel}. The respective number
densities are $\rho_C(\rv)$, $\rho_P(\rv)$, and $\rho_S(\rv)$,
where $\rv$ is the spatial coordinate. The primary solvent is
regarded as a homogeneous background for the polymer and is not
explicitly included. All particles experience only pairwise
interactions, $V_{ij}(r)$, $i,j=C,P,S$, where $r$ is the
separation distance between particle centers.  Colloids behave as
hard spheres: $V_{CC}(r)=\infty$, if $r<2R_C$, and zero otherwise.
Colloids and polymers interact as hard bodies via
$V_{CP}(r)=\infty$, if $r<R_C+R_P$, and zero otherwise, and both
exclude cosolvent particles: $V_{CS}(r)=\infty$, if $r<R_C$,
$V_{PS}(r)=\infty$, if $r<R_P$, and zero otherwise. The polymers
and cosolvent particles behave as ideal gases: $V_{PP}(r)=0$,
$V_{SS}(r)=0$, for all $r$. In essence, this is the AO model with
additional point particles that cannot penetrate either colloids
or polymers.

We denote the sphere diameters by $\sigma_C=2R_C$ and
$\sigma_P=2R_P$, the bulk packing fractions by $\eta_C=4\pi R_C^3
\rho_C/3$ and $\eta_P=4\pi R_P^3 \rho_P/3$, and define a dimensionless
solvent bulk density $\rs=\rho_S \sigma_P^3.$ The polymer-colloid size
ratio, $q=\sigma_P/\sigma_C$, is regarded as a control parameter.

\section{Density functional theory}
\label{SECtheory} We develop a geometry-based DF theory for the
excess Helmholtz free energy of the model system, expressed as an
integral over an excess free energy density,
\begin{equation}
\Fexc [ \rho_C, \rho_P, \rho_S ] = \kt \int {\rm d}^3 x\ \phifmt{}
\left( \{\n{\nu}{i}\} \right), \label{EQfexc}
\end{equation}
where $\kb$ is Boltzmann's constant, $T$ is absolute temperature,
and the (local) reduced excess free energy density, $\phifmt{}$,
is a simple function (not a functional) of weighted densities,
$\n{\nu}{i}$. The weighted densities are smoothed averages of the
possibly highly inhomogeneous density profiles, $\rho_i(\rv)$,
expressed as convolutions,
\begin{equation}
  \n{\fumindex}{i}(\rv) = \rho_i(\rv) * \w{\fumindex}{i}(\rv)
  = \int{\rm d}\rv'\ \rho_i(\rv') w_{\nu}^i(\rv-\rv'),
    \label{EQnfirst}
\end{equation}
with respect to weight functions, $\w{\fumindex}{i}(\rv)$, where
$i=C,P,S$ and $\nu=$0,1,2,3,v1,v2,m2. The usual weight
functions~\cite{Rosenfeld89,tarazona00} are
\begin{equation}
\w{2}{i}( \rv ) = \delta( R_i - r), \hspace{3mm}
\w{3}{i}( \rv ) = \theta( R_i - r),  
\label{EQwsfirst} 
\end{equation}
\vspace{-0.5cm}
\begin{equation}
\wv{2}{i}( \rv ) = \w{2}{i}(\rv) \, \frac{\rv}{r},  \hspace{3mm}
\wm{2}{i}( \rv ) =  \w{2}{i}(\rv) \left( \frac{\rv\rv}{r^2} - 
\frac{\mymat{1}}{3} \right),
\label{EQwssecond}
\end{equation}
where $r=|\rv|$, $\delta(r)$ is the Dirac distribution, 
$\theta(r)$ is the step function, and $\mymat{1}$ is the identity matrix.
Further linearly dependent weight functions are $\w{1}{i}(\rv) =
\w{2}{i}(\rv)/(4 \pi R), \wv{1}{i}(\rv) = \wv{2}{i}(\rv)/(4 \pi
R)$, and $\w{0}{i}(\rv) = \w{1}{i}(\rv)/R$. The weight functions
for $\nu=3,2,1,0$ represent geometrical measures of the particles
in terms of volume, surface area, integral mean curvature, and
Euler characteristic, respectively~\cite{Rosenfeld89}. Note that
the weight functions differ in tensorial rank: $\w{0}{i}$,
$\w{1}{i}$, $\w{2}{i}$, and $\w{3}{i}$ are scalars, $\wv{1}{i}$
and $\wv{2}{i}$ are vectors, and $\wm{2}{i}$ is a (traceless)
matrix.

The excess free energy density can be expressed in the general
form
\begin{equation}
 \Phi = \Phi_{C} + \Phi_{CP} +  \Phi_{CS} + \Phi_{CPS},
\end{equation}
where the four contributions have forms motivated by consideration
of the appropriate exact zero-dimensional limits. The colloid
contribution, $\Phi_{C}$, is the same as that for the pure
hard-sphere (HS) system~\cite{Rosenfeld89,tarazona00}:
\begin{eqnarray}
\phifmt{C} &=&
  -\n{0}{C} \ln (1 - \n{3}{C})+
  \frac{\n{1}{C}\,\n{2}{C} -
  \nv{1}{C} \cdot \nv{2}{C}}{1 -
  \n{3}{C}} \nonumber\\
& &+
\left[{\frac{1}{3}{\left(\n{2}{C}\right)^3}} -
  \n{2}{C}\,{\left(\nv{2}{C}\right)^2} +
  \frac{3}{2}\left(\nv{2}{C} \cdot \nmtl{2}{C} \cdot \nv{2}{C}
\right.\right.\nonumber\\
&&  \left.\left.\phantom{{\left(\n{2}{C}\right)^3}/3}\left.\hspace{-8mm}
          -\,3\det \nmtl{2}{C}\right)
      \right]\right/[8\pi(1-\n{3}{C})^2].\label{EQphic}
\end{eqnarray}
The colloid-polymer interaction contribution, $\phifmt{CP}$, is
the same as in the pure AO case~\cite{schmidt00cip},
\begin{equation}
\phifmt{CP} =
 \sum_\nu \frac{\partial \phifmt{C}}{\partial \n{\nu}{C}}
 \n{\nu}{P},
\end{equation}
while the colloid-solvent interaction
contribution~\cite{schmidt01rsf} is
\begin{equation}
\phifmt{CS} = -\n{0}{S} \ln (1 - \n{3}{C})\label{EQphics}.
\end{equation}
Finally, in order to model the WR-type interaction between
polymers and cosolvent particles in the presence of the 
colloidal spheres, we assume
\begin{equation}
\phifmt{CPS} =
  \frac{ \n{0}{S} \n{3}{P} }{1-\n{3}{C}}\label{EQphipn},
  \label{EQphicps}
\end{equation}
which takes into account the volume excluded to the polymer and
cosolvent by the colloids.

It is instructive to compare the current theory to geometry-based
DF theories previously formulated for two related ternary model
systems. One starting point is a ternary AO model that combines a
binary HS mixture and one polymer species~\cite{schmidt02cip}. 
Letting the radius of the smaller HS component
go to zero, one obtains the cosolvent species. The other starting
point is a recently-introduced model~\cite{schmidt02cpn} for a
ternary mixture of colloids, polymers and hard vanishingly thin
needles of length $L$, where the needles are ideal amongst
themselves but cannot penetrate the polymers (hard-core
interaction).  In the limit $L\to 0$, the needles become identical
to the cosolvent particles. We have explicitly checked that the DF
theories for both systems reduce to the theory described above,
demonstrating the internal consistency of the geometry-based
approach.

\section{Results and Discussion}
\label{SECresults}
\subsection{Bulk Limit}

For bulk fluid phases the density profiles are homogeneous:
$\rho_i(\rv) =$ const. In this case, the integrations in 
Eq.\ (\ref{EQnfirst}) are trivial, and simple expressions for the
weighted densities can be obtained. Inserting these expressions
into the excess free energy density 
[Eqs.\ (\ref{EQphic})-(\ref{EQphicps})] yields the bulk excess 
free energy in analytic form. 
The HS contribution, which is equal to the
Percus-Yevick compressibility (and scaled-particle) result, is
given as
\begin{equation}
  \Phi_{C} = \frac{3 \eta_C
    [3 \eta_C (2-\eta_C) - 2(1-\eta_C)^2\ln(1-\eta_C)]}
  {8 \pi R_C^3 (1-\eta_C)^2}.
\end{equation}
The colloid-polymer contribution is equal to that predicted by
free volume theory~\cite{lekkerkerker92}, and subsequently
rederived by DFT~\cite{schmidt00cip}:
\begin{eqnarray}
  \Phi_{CP} &=&
    \frac{\eta_P/(8\pi R_P^3)}{(1-\eta_C)^3}
    \left\{
    3 q \eta_C \left[ 6(1-\eta_C)^2 
\right.\right.
\nonumber\\
     &+&\left. 3q(2-\eta_C -\eta_C^2)
     +2q^2(1+\eta_C+\eta_C^2)  \right] 
\nonumber\\
     &-&\left. 6(1-\eta_C)^3 \ln (1 - \eta_C)
    \right\}.
\end{eqnarray}
This contribution is linear in the polymer density and has a form that 
arises, as in the original free volume theory~\cite{lekkerkerker92}, 
from treating the polymers as an ideal gas occupying the free volume 
between the colloids.  The colloid-cosolvent contribution is given by
\begin{equation}
  \Phi_{CS} = -\rho_S\ln(1-\eta_C).
\end{equation}
This contribution can be similarly interpreted as the free energy 
of an ideal gas in the free volume of the colloids.  In this case, 
however, the ideal gas consists of point-like cosolvent particles,
considerably simplifying the analytical form of the free volume. 
In fact, by letting $q\to 0$ in Eq.~(11), and identifying species 
$P$ and $S$, $\Phi_{CP}$ reduces to $\Phi_{CS}$.
The remaining contribution couples the densities of all three species,
and is given by
\begin{equation}
  \Phi_{CPS} = \frac{\rho_S \eta_P}{1-\eta_C}.
\end{equation}
In the absence of colloids ($\eta_C=0$), this is equivalent to the
mean-field free energy of the WR-model.  Eq.~(13) is a non-trivial
generalization thereof to the case of non-vanishing $\eta_C$.
For completeness, the reduced ideal-gas free energy is
\begin{equation}
  \Phi_{\rm id} = \sum_{i=C,P,S} \rho_i
  [\ln(\rho_i \Lambda_i^3) -1],
\end{equation}
where the $\Lambda_i$ are (irrelevant) thermal wavelengths of
species $i$.  This puts us in a position to obtain the reduced
total free energy density, $\Phi_{\rm tot}=\Phi_{\rm id}+\Phi$, of
any given fluid state characterized by the bulk densities of the
three components and the size ratio $q$.

\subsection{Phase Diagrams}

The conditions for phase coexistence are equality of the total
pressures, $p_{\rm tot}$, and of the chemical potentials, $\mu_i$,
in the coexisting phases. For phase equilibrium between phases I
and II,
$p_{\rm tot}^{\rm I} = p_{\rm tot}^{\rm II}$ and $\mu_i^{\rm I} =
\mu_i^{\rm II}, i=C,P,S$, yielding four equations for six unknowns
(two state-points, each characterized by three densities). In our
case, a set of analytical expressions is obtained from
\begin{equation}
\frac{p_{\rm tot}}{k_BT}=-\Phi_{\rm tot}+\sum_{i=C,P,S} \rho_i
~\frac{\partial \Phi_{\rm tot}}{\partial \rho_i}
\end{equation}
and
\begin{equation}
\mu_i=k_B T~\frac{\partial \Phi_{\rm tot}}{\partial \rho_i},
\end{equation}
the numerical solution of which is straightforward.

In order to graphically represent the ternary phase diagrams, we
choose the system reduced densities, $\ec,\ep$, and $\rs$ as basic
variables. For given $q$, these span a three-dimensional (3d)
phase space. Each point in this space corresponds to a possible
bulk state. Two-phase coexistence is indicated by a pair of points
joined by a straight tie-line. We imagine controlling the system
directly with $\ec$ and $\ep$, but indirectly via coupling to a
cosolvent reservoir, whose chemical potential, $\mu_S$, tunes 
the solvent quality.  Note that, because the cosolvent is treated 
as an ideal gas, the reservoir's density is simply proportional to 
its activity.  Thus, the reduced density, $\rsr= \exp(\mu_S/k_BT)$, 
may be equivalently taken as a control parameter, which is
equal in coexisting phases.
To make contact with Flory-Huggins theory, we are implicitly 
considering here the case in which the Flory interaction parameter, 
$\chi$, falls in the range $0.5 < \chi < 1$, corresponding to a 
negative excluded-volume parameter, $v \propto (1-2\chi)$.

\begin{figure}
  \begin{center}
    \includegraphics[width=\mypicwidth,angle=-90]{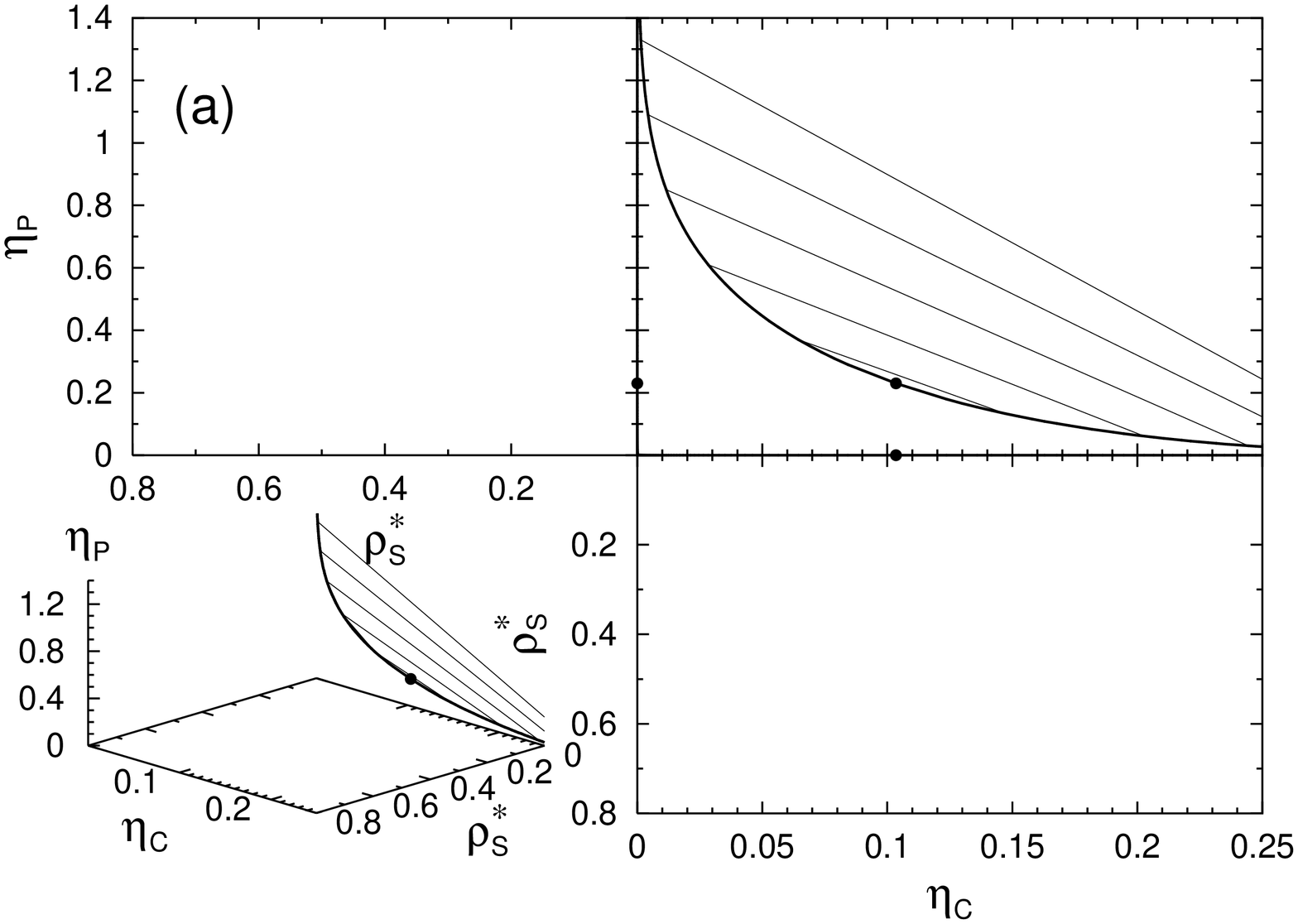}
    \includegraphics[width=\mypicwidth,angle=-90]{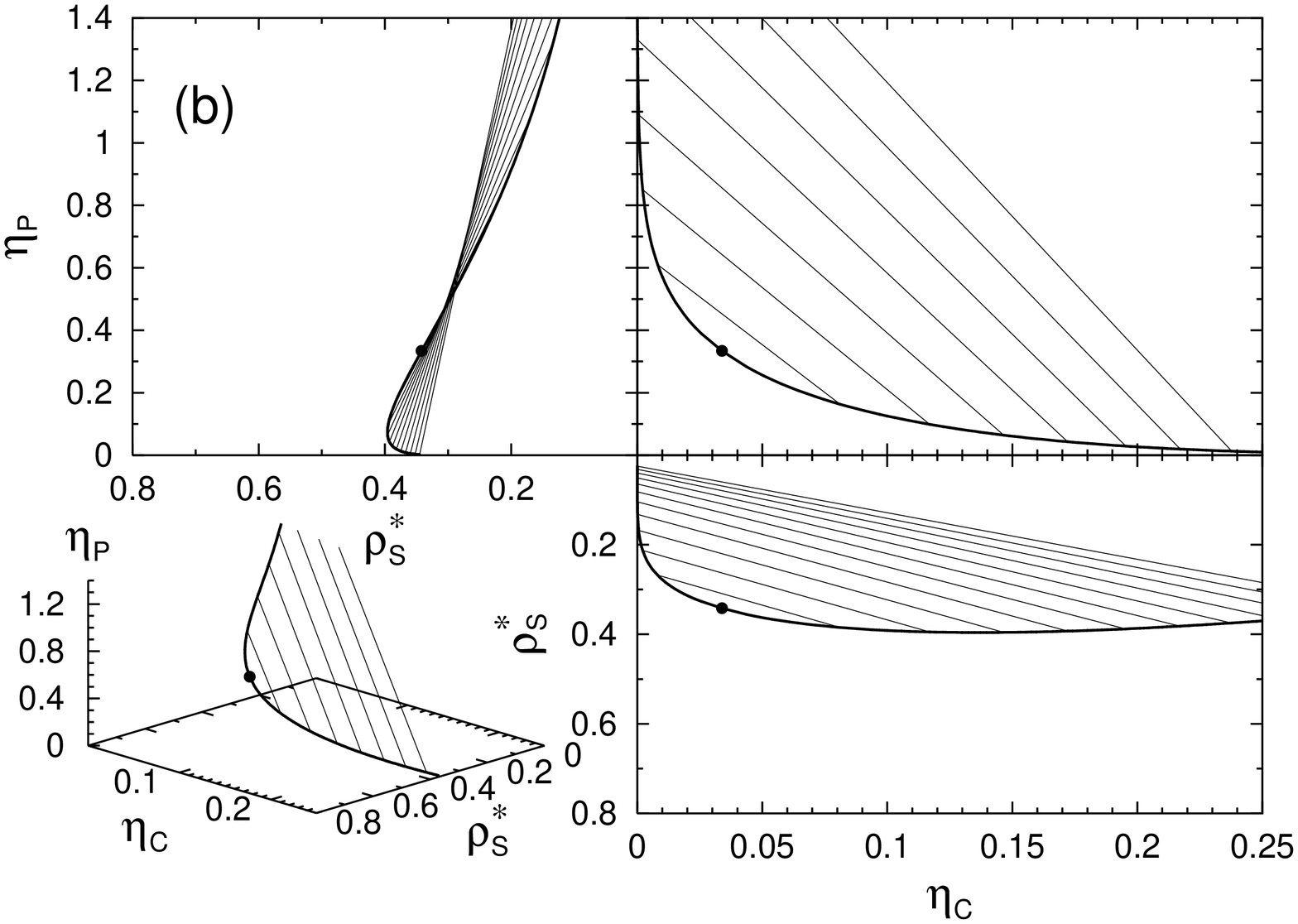}
    \includegraphics[width=\mypicwidth,angle=-90]{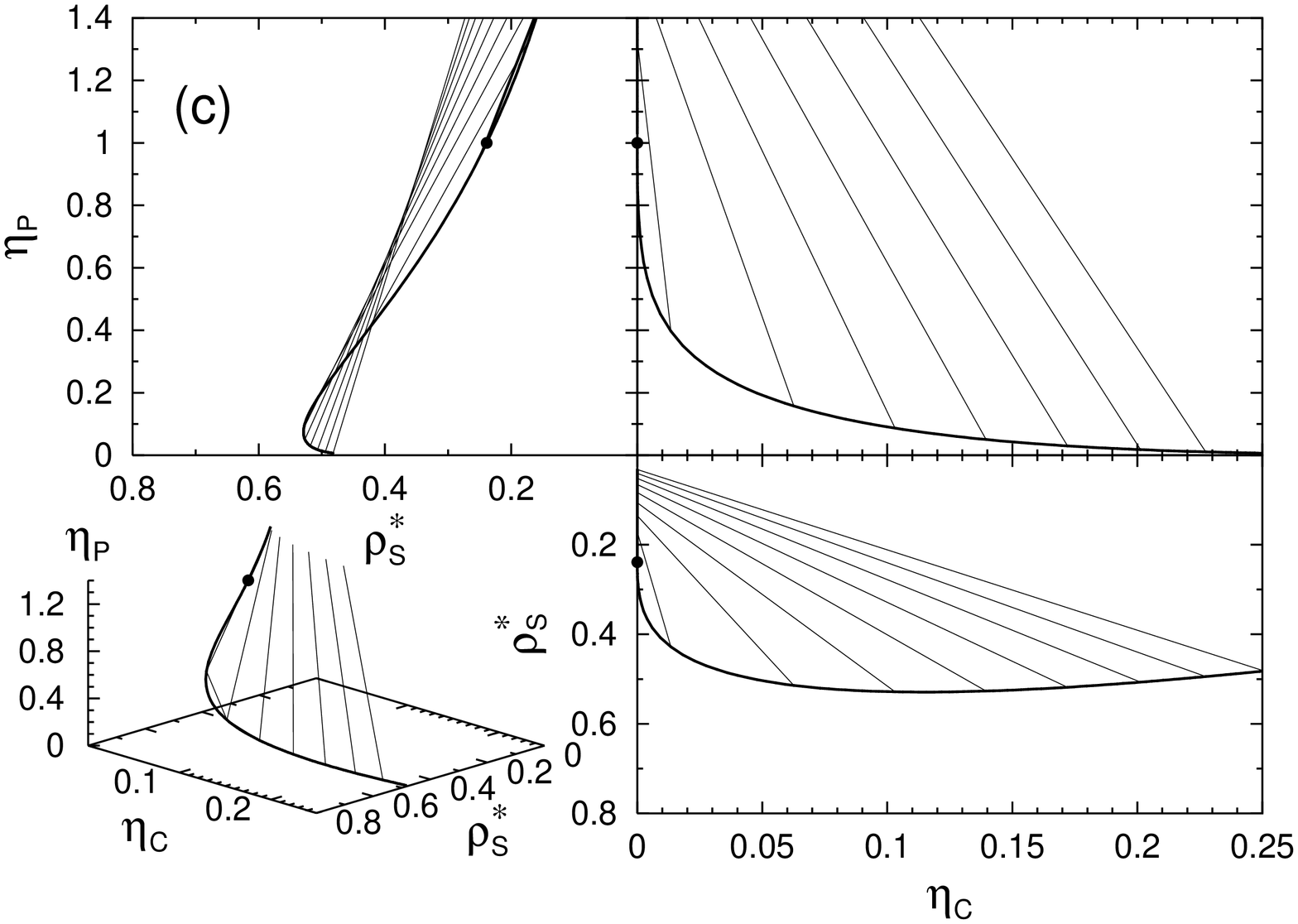}
    \includegraphics[width=\mypicwidth,angle=-90]{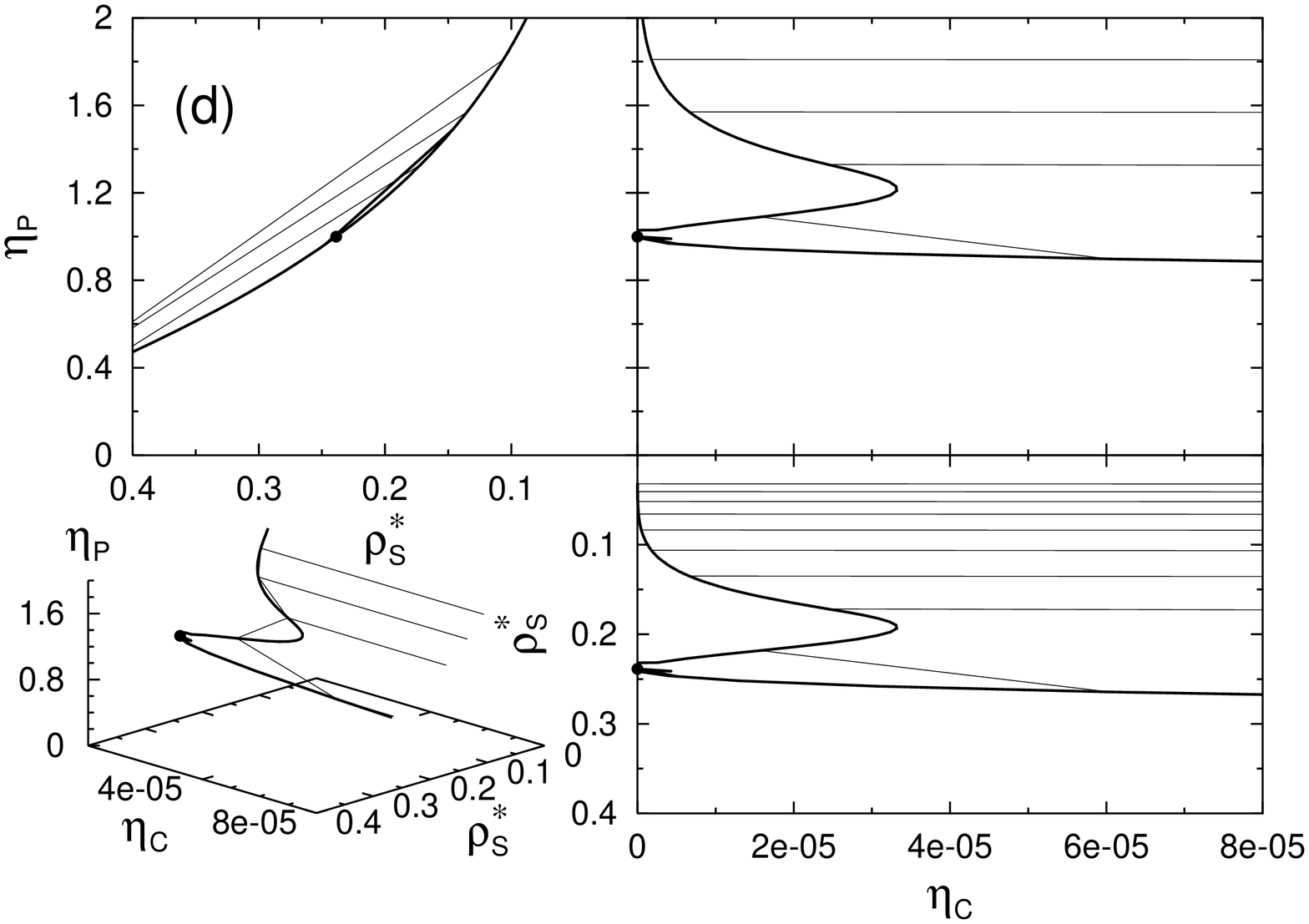} \\
    \vspace{1cm}
    \caption{Demixing phase diagram of the model ternary
    colloid-polymer-solvent mixture for $\sigma_C=\sigma_P$
    and $\rsr=0$ (a), 0.5 (b), and 0.64894 (c). The latter case
    is shown also on a larger scale (d).}
    \label{FIGps1} 
   \end{center}
\end{figure}

We initially consider colloids and polymers of equal size
($\sigma_C=\sigma_P$). For this case, Fig.~\ref{FIGps1} shows
projections of constant-$\rsr$ surfaces onto the three sides of
the coordinate system, namely the $\ec-\rs$, $\ec-\ep$, and
$\ep-\rs$ planes, as well as a perspective 3d view. For reference,
the phase diagram without cosolvent is shown in
Fig.~\ref{FIGps1}a. This is identical to the common free volume
demixing curve of the AO model~\cite{lekkerkerker92,schmidt00cip}.
For $\rsr=0$, in which case $\rs=0$,
the $\ec-\rs$ and $\ep-\rs$ planes are inaccessible, {\it i.e.},
all accessible states lie completely within the $\ec-\ep$ plane.  Upon
increasing the cosolvent reservoir density to $\rsr=0.5$, and thus
worsening the solvent quality, the demixed region grows, as seen
in Fig.~\ref{FIGps1}b. The critical point shifts towards lower
$\ec$ and higher $\ep$, the tie lines become steeper, and the area
beneath the colloid-polymer binodal in the $\ec-\ep$ plane (a
measure of miscibility) decreases.
As a physical interpretation of the results, one can imagine the
polymer spheres as tending to merge (overlap) to avoid contact
with the solvent. The resulting polymer ``dimers,'' ``trimers,''
etc., act as larger depleting agents, increasing the range of the
effective depletion potential between colloids. At the same time,
the lower effective concentration of depletants reduces the
osmotic pressure and thus the depth of the potential. Comparing
the phase diagrams for different cosolvent reservoir densities, we
can conclude that the net effect of merging polymers is to
increase the integrated strength of the depletion potential and
thus to promote demixing. 

Eventually, at $\rsr=0.64894$, the colloid-polymer critical point
meets the $\ep-\rs$ plane (where $\ec=0$), as seen in
Figs.~\ref{FIGps1}c and (on a larger scale) \ref{FIGps1}d.
Polymers and cosolvent here begin to demix already in the absence
of colloids (the critical point of the WR model). For still higher
cosolvent reservoir densities (beyond the WR critical point), the
critical point vanishes from the phase diagram and a
polymer-cosolvent miscibility gap opens up at $\ec=0$. It is
tempting to interpret this demixing as aggregation of the polymer
spheres, although it must be emphasized that the WR model can only
crudely describe polymer aggregation.

Another intriguing prediction is the reentrant colloid-polymer
mixing evident in Fig.~\ref{FIGps1}d. For sufficiently low colloid
concentrations and high cosolvent reservoir densities (poor
solvent), colloids and polymers initially demix with increasing
$\ep$. Upon increasing $\ep$ further, miscibility returns over a
small range before demixing again occurs at higher $\ep$. Such a
phenomenon could conceivably result from the complex interplay
between range and depth of the depletion potential arising from
solvent-induced overlap of polymers.

\begin{figure}
  \begin{center}
    \includegraphics[width=\mypicwidth,angle=-90]{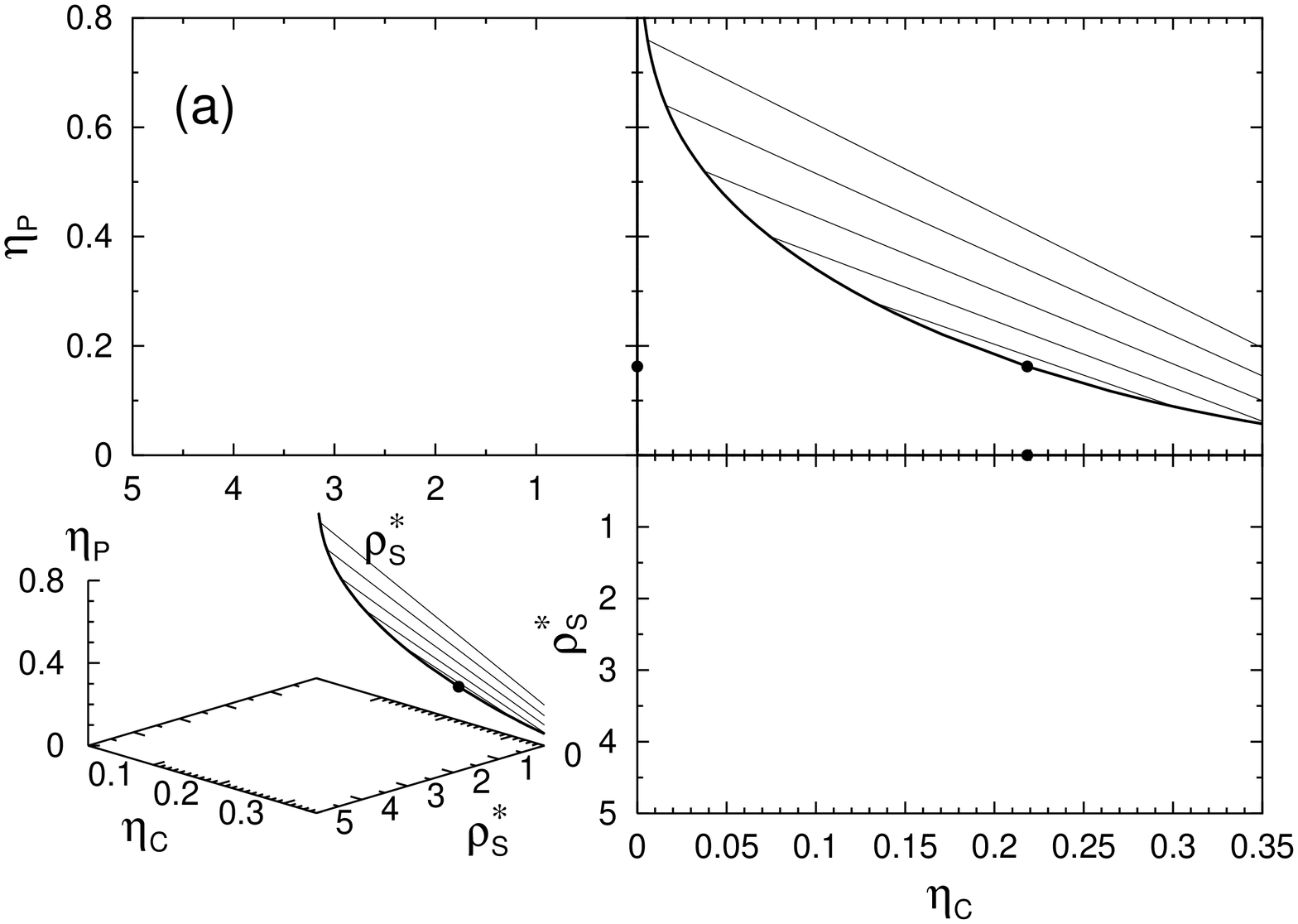}
    \includegraphics[width=\mypicwidth,angle=-90]{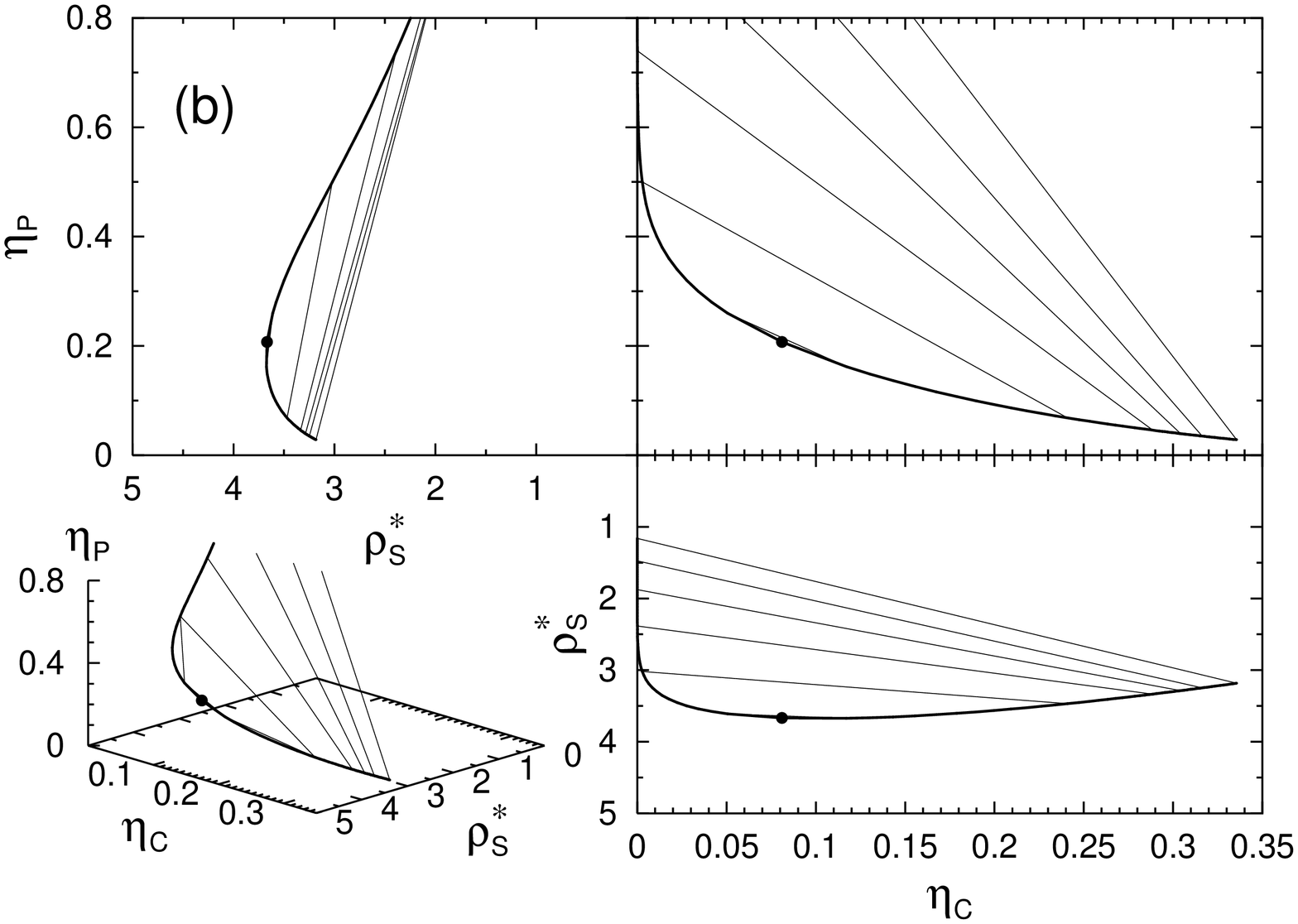} 
    \vspace{1cm}
    \caption{Same as Fig.~\ref{FIGps1}, but for $\sigma_C=2\sigma_P$
    and $\rsr=0$ (a) and 0.5 (b).}
    \label{FIGps2} 
   \end{center}
\end{figure}

For smaller polymer-to-colloid size ratios, the above scenario
persists. Figure~\ref{FIGps2} shows qualitatively similar results
for $q=0.5$ and cosolvent reservoir densities $\rsr=0$
(Fig.~\ref{FIGps2}a) and 0.5 (Fig.~\ref{FIGps2}b).

\vspace{1cm}

\section{Conclusions}
\label{SECdiscussion}

In summary, we have investigated the bulk fluid demixing behavior
of model mixtures of colloids and nonadsorbing polymers in poor
solvents. Our model combines the Asakura-Oosawa model of 
hard-sphere colloids plus ideal penetrable-sphere polymers with
a binary solvent model. The solvent comprises a primary theta-solvent 
and a cosolvent of point particles that are excluded from both colloids
and polymers. Cosolvent exclusion energetically favors overlapping 
configurations of polymers.  Although somewhat idealized, the 
model exhibits the essential feature of solvent-induced effective 
attraction between polymers, mimicking the effect of a poor solvent. 

To study the equilibrium phase behavior of this model, we have derived 
a geometry-based density functional theory that combines elements of
previous theories for the AO and Widom-Rowlinson models. Applying
the theory to bulk fluid phases, we have calculated phase diagrams
for cosolvent densities spanning a range from theta-solvent to
poor solvent. With increasing cosolvent concentration (worsening
solvent quality), the predicted colloid-polymer binodal shifts to
lower colloid concentrations, destabilizing the mixed phase.
Beyond a threshold cosolvent concentration, a reentrant
colloid-polymer demixing transition is predicted at low colloid
concentrations.

Predictions of the theory could be tested by comparison with 
simulations of the model. 
Qualitative comparison with experiment also may be possible, 
but would require a relation between the cosolvent concentration 
(as a measure of solvent quality) and the Flory interaction parameter. 
In principle, such a relation could be established by calculating 
the effective second virial coefficient of the polymer in the 
polymer-cosolvent subsystem.

Although here we have approximated the polymers as mutually
noninteracting, their effective attractions being driven only
by cosolvent exclusion, future work should include non-ideality 
between polymers, arising fundamentally from excluded-volume
repulsion between polymer segments. For this purpose, a reasonable
model is an effective-sphere description based on a repulsive,
penetrable pair interaction (finite at the origin), {\it e.g.}, 
of step-function or Gaussian shape~\cite{louis00}. The competition
between such intrinsic repulsion and the solvent-induced
attraction considered in this work is likely to produce rich
phase behavior. As a further outlook, our approach also could be
applied to effects of solvent quality on polymer brushes adsorbed
onto surfaces of colloidal particles.


\bibliographystyle{apsrev}
\bibliography{cpps}

\begin{thebibliography}{22}
\expandafter\ifx\csname natexlab\endcsname\relax\def\natexlab#1{#1}\fi
\expandafter\ifx\csname bibnamefont\endcsname\relax
  \def\bibnamefont#1{#1}\fi
\expandafter\ifx\csname bibfnamefont\endcsname\relax
  \def\bibfnamefont#1{#1}\fi
\expandafter\ifx\csname citenamefont\endcsname\relax
  \def\citenamefont#1{#1}\fi
\expandafter\ifx\csname url\endcsname\relax
  \def\url#1{\texttt{#1}}\fi
\expandafter\ifx\csname urlprefix\endcsname\relax\def\urlprefix{URL }\fi
\providecommand{\bibinfo}[2]{#2}
\providecommand{\eprint}[2][]{\url{#2}}

\bibitem[{\citenamefont{Flory}(1969)}]{flory69}
\bibinfo{author}{\bibfnamefont{P.~J.} \bibnamefont{Flory}},
  \emph{\bibinfo{title}{Statistical Mechanics of Chain Molecules}}
  (\bibinfo{publisher}{Interscience Publishers}, \bibinfo{address}{New York},
  \bibinfo{year}{1969}).

\bibitem[{\citenamefont{de~Gennes}(1979)}]{deGennes79}
\bibinfo{author}{\bibfnamefont{P.-G.} \bibnamefont{de~Gennes}},
  \emph{\bibinfo{title}{Scaling Concepts in Polymer Physics}}
  (\bibinfo{publisher}{Cornell University Press}, \bibinfo{address}{Ithaca,
  N.Y.}, \bibinfo{year}{1979}).

\bibitem[{\citenamefont{Flory}(1971)}]{flory71}
\bibinfo{author}{\bibfnamefont{P.~J.} \bibnamefont{Flory}},
  \emph{\bibinfo{title}{Principles of Polymer Chemistry}}
  (\bibinfo{publisher}{Cornell University Press}, \bibinfo{address}{Ithaca,
  N.Y.}, \bibinfo{year}{1971}).

\bibitem[{\citenamefont{Asakura and Oosawa}(1954)}]{asakura54}
\bibinfo{author}{\bibfnamefont{S.}~\bibnamefont{Asakura}} \bibnamefont{and}
  \bibinfo{author}{\bibfnamefont{F.}~\bibnamefont{Oosawa}},
  \bibinfo{journal}{J. Chem. Phys.} \textbf{\bibinfo{volume}{22}},
  \bibinfo{pages}{1255} (\bibinfo{year}{1954}).

\bibitem[{\citenamefont{Vrij}(1976)}]{vrij76}
\bibinfo{author}{\bibfnamefont{A.}~\bibnamefont{Vrij}}, \bibinfo{journal}{Pure
  and Appl. Chem.} \textbf{\bibinfo{volume}{48}}, \bibinfo{pages}{471}
  (\bibinfo{year}{1976}).

\bibitem[{\citenamefont{Gast et~al.}(1983)\citenamefont{Gast, Hall, and
  Russell}}]{gast83}
\bibinfo{author}{\bibfnamefont{A.~P.} \bibnamefont{Gast}},
  \bibinfo{author}{\bibfnamefont{C.~K.} \bibnamefont{Hall}}, \bibnamefont{and}
  \bibinfo{author}{\bibfnamefont{W.~B.} \bibnamefont{Russell}},
  \bibinfo{journal}{J. Coll. Int. Sci.} \textbf{\bibinfo{volume}{96}},
  \bibinfo{pages}{251} (\bibinfo{year}{1983}).

\bibitem[{\citenamefont{Lekkerkerker et~al.}(1992)\citenamefont{Lekkerkerker,
  Poon, Pusey, Stroobants, and Warren}}]{lekkerkerker92}
\bibinfo{author}{\bibfnamefont{H.~N.~W.} \bibnamefont{Lekkerkerker}},
  \bibinfo{author}{\bibfnamefont{W.~C.~K.} \bibnamefont{Poon}},
  \bibinfo{author}{\bibfnamefont{P.~N.} \bibnamefont{Pusey}},
  \bibinfo{author}{\bibfnamefont{A.}~\bibnamefont{Stroobants}},
  \bibnamefont{and} \bibinfo{author}{\bibfnamefont{P.~B.}
  \bibnamefont{Warren}}, \bibinfo{journal}{Europhys. Lett.}
  \textbf{\bibinfo{volume}{20}}, \bibinfo{pages}{559} (\bibinfo{year}{1992}).

\bibitem[{\citenamefont{Schmidt et~al.}(2000)\citenamefont{Schmidt, L{\"o}wen,
  Brader, and Evans}}]{schmidt00cip}
\bibinfo{author}{\bibfnamefont{M.}~\bibnamefont{Schmidt}},
  \bibinfo{author}{\bibfnamefont{H.}~\bibnamefont{L{\"o}wen}},
  \bibinfo{author}{\bibfnamefont{J.~M.} \bibnamefont{Brader}},
  \bibnamefont{and} \bibinfo{author}{\bibfnamefont{R.}~\bibnamefont{Evans}},
  \bibinfo{journal}{Phys. Rev. Lett.} \textbf{\bibinfo{volume}{85}},
  \bibinfo{pages}{1934} (\bibinfo{year}{2000}).

\bibitem[{\citenamefont{Dijkstra et~al.}(1999)\citenamefont{Dijkstra, Brader,
  and Evans}}]{dijkstra99}
\bibinfo{author}{\bibfnamefont{M.}~\bibnamefont{Dijkstra}},
  \bibinfo{author}{\bibfnamefont{J.~M.} \bibnamefont{Brader}},
  \bibnamefont{and} \bibinfo{author}{\bibfnamefont{R.}~\bibnamefont{Evans}},
  \bibinfo{journal}{J. Phys.: Condens. Matter} \textbf{\bibinfo{volume}{11}},
  \bibinfo{pages}{10079} (\bibinfo{year}{1999}).

\bibitem[{\citenamefont{Warren et~al.}(1995)\citenamefont{Warren, Ilett, and
  Poon}}]{warren95}
\bibinfo{author}{\bibfnamefont{P.~B.} \bibnamefont{Warren}},
  \bibinfo{author}{\bibfnamefont{S.~M.} \bibnamefont{Ilett}}, \bibnamefont{and}
  \bibinfo{author}{\bibfnamefont{W.~C.~K.} \bibnamefont{Poon}},
  \bibinfo{journal}{Phys. Rev. E} \textbf{\bibinfo{volume}{52}},
  \bibinfo{pages}{5205} (\bibinfo{year}{1995}).

\bibitem[{\citenamefont{Schmidt
  et~al.}(2002{\natexlab{a}})\citenamefont{Schmidt, Denton, and
  Brader}}]{schmidt02intpol}
\bibinfo{author}{\bibfnamefont{M.}~\bibnamefont{Schmidt}},
  \bibinfo{author}{\bibfnamefont{A.~R.} \bibnamefont{Denton}},
  \bibnamefont{and} \bibinfo{author}{\bibfnamefont{J.~M.} \bibnamefont{Brader}}
  (\bibinfo{year}{2002}{\natexlab{a}}), \bibinfo{note}{unpublished}.

\bibitem[{\citenamefont{Louis et~al.}(2000)\citenamefont{Louis, Bolhuis,
  Hansen, and Meijer}}]{louis00}
\bibinfo{author}{\bibfnamefont{A.~A.} \bibnamefont{Louis}},
  \bibinfo{author}{\bibfnamefont{P.~G.} \bibnamefont{Bolhuis}},
  \bibinfo{author}{\bibfnamefont{J.~P.} \bibnamefont{Hansen}},
  \bibnamefont{and} \bibinfo{author}{\bibfnamefont{E.~J.}
  \bibnamefont{Meijer}}, \bibinfo{journal}{Phys. Rev. Lett.}
  \textbf{\bibinfo{volume}{85}}, \bibinfo{pages}{2522} (\bibinfo{year}{2000}).

\bibitem[{\citenamefont{{P. G. Bolhuis, A. A. Louis, J. P. Hansen, and E. J.
  Meijer}}(2001)}]{bolhuis01jcp}
\bibinfo{author}{\bibnamefont{{P. G. Bolhuis, A. A. Louis, J. P. Hansen, and E.
  J. Meijer}}}, \bibinfo{journal}{J. Chem. Phys.}
  \textbf{\bibinfo{volume}{114}}, \bibinfo{pages}{4296} (\bibinfo{year}{2001}).

\bibitem[{\citenamefont{{P. G. Bolhuis, A. A. Louis, and J. P.
  Hansen}}(2001)}]{bolhuis01pre}
\bibinfo{author}{\bibnamefont{{P. G. Bolhuis, A. A. Louis, and J. P. Hansen}}},
  \bibinfo{journal}{Phys. Rev. E} \textbf{\bibinfo{volume}{64}},
  \bibinfo{pages}{021801} (\bibinfo{year}{2001}).

\bibitem[{\citenamefont{Ramakrishnan et~al.}(2002)\citenamefont{Ramakrishnan,
  Fuchs, Schweizer, and Zukoski}}]{ramakrishnan02}
\bibinfo{author}{\bibfnamefont{S.}~\bibnamefont{Ramakrishnan}},
  \bibinfo{author}{\bibfnamefont{M.}~\bibnamefont{Fuchs}},
  \bibinfo{author}{\bibfnamefont{K.~S.} \bibnamefont{Schweizer}},
  \bibnamefont{and} \bibinfo{author}{\bibfnamefont{C.~F.}
  \bibnamefont{Zukoski}}, \bibinfo{journal}{J. Chem. Phys.}
  \textbf{\bibinfo{volume}{116}}, \bibinfo{pages}{2201} (\bibinfo{year}{2002}).

\bibitem[{\citenamefont{Widom and Rowlinson}(1970)}]{widom70}
\bibinfo{author}{\bibfnamefont{B.}~\bibnamefont{Widom}} \bibnamefont{and}
  \bibinfo{author}{\bibfnamefont{J.~S.} \bibnamefont{Rowlinson}},
  \bibinfo{journal}{J. Chem. Phys.} \textbf{\bibinfo{volume}{52}},
  \bibinfo{pages}{1670} (\bibinfo{year}{1970}).

\bibitem[{\citenamefont{Rowlinson and Widom}(1982)}]{rowlinson82}
\bibinfo{author}{\bibfnamefont{J.~S.} \bibnamefont{Rowlinson}}
  \bibnamefont{and} \bibinfo{author}{\bibfnamefont{B.}~\bibnamefont{Widom}},
  \emph{\bibinfo{title}{Molecular Theory of Capillarity}}
  (\bibinfo{publisher}{Clarendon Press}, \bibinfo{address}{Oxford},
  \bibinfo{year}{1982}).

\bibitem[{\citenamefont{Rosenfeld}(1989)}]{Rosenfeld89}
\bibinfo{author}{\bibfnamefont{Y.}~\bibnamefont{Rosenfeld}},
  \bibinfo{journal}{Phys. Rev. Lett.} \textbf{\bibinfo{volume}{63}},
  \bibinfo{pages}{980} (\bibinfo{year}{1989}).

\bibitem[{\citenamefont{Tarazona}(2000)}]{tarazona00}
\bibinfo{author}{\bibfnamefont{P.}~\bibnamefont{Tarazona}},
  \bibinfo{journal}{Phys. Rev. Lett.} \textbf{\bibinfo{volume}{84}},
  \bibinfo{pages}{694} (\bibinfo{year}{2000}).

\bibitem[{\citenamefont{Schmidt}(2001)}]{schmidt01rsf}
\bibinfo{author}{\bibfnamefont{M.}~\bibnamefont{Schmidt}},
  \bibinfo{journal}{Phys. Rev. E} \textbf{\bibinfo{volume}{63}},
  \bibinfo{pages}{050201(R)} (\bibinfo{year}{2001}).

\bibitem[{\citenamefont{Schmidt
  et~al.}(2002{\natexlab{b}})\citenamefont{Schmidt, L{\"o}wen, Brader, and
  Evans}}]{schmidt02cip}
\bibinfo{author}{\bibfnamefont{M.}~\bibnamefont{Schmidt}},
  \bibinfo{author}{\bibfnamefont{H.}~\bibnamefont{L{\"o}wen}},
  \bibinfo{author}{\bibfnamefont{J.~M.} \bibnamefont{Brader}},
  \bibnamefont{and} \bibinfo{author}{\bibfnamefont{R.}~\bibnamefont{Evans}}
  (\bibinfo{year}{2002}{\natexlab{b}}), \bibinfo{note}{submitted to J. Phys:
  Condens. Matter}.

\bibitem[{\citenamefont{Schmidt and Denton}(2002)}]{schmidt02cpn}
\bibinfo{author}{\bibfnamefont{M.}~\bibnamefont{Schmidt}} \bibnamefont{and}
  \bibinfo{author}{\bibfnamefont{A.~R.} \bibnamefont{Denton}},
  \bibinfo{journal}{Phys. Rev. E} \textbf{\bibinfo{volume}{65}},
  \bibinfo{pages}{021508} (\bibinfo{year}{2002}).

\end{thebibliography}

\end{document}